\documentclass[aps,prb,twocolumn,superscriptaddress,nofootinbib,longbibliography]{revtex4-2}
\usepackage{mathrsfs,color}
\usepackage{booktabs}
\usepackage{float}
\usepackage{tabularx}
\usepackage{epsfig,float}
\usepackage{graphicx}
\usepackage{amsfonts}
\usepackage[figuresright]{rotating}
\usepackage{amssymb}
\usepackage{amsmath}
\usepackage{dcolumn}
\usepackage{bm}
\usepackage{comment}
\usepackage{xcolor}
\usepackage{color}
\usepackage{braket}
\usepackage{units}
\usepackage{xspace}

\usepackage{bbold}
\definecolor{mypink3}{cmyk}{0, 0.7808, 0.4429, 0.1412}
\definecolor{mypink1}{rgb}{0.858, 0.188, 0.478}
\definecolor{mypink2}{RGB}{219, 48, 122}
\usepackage[colorlinks=true, allcolors=mypink1]{hyperref}
\usepackage[fleqn]{mathtools}

\usepackage[shortlabels]{enumitem}

\begin{document}

\title{Fragmentation in Non-Hermitian Many-Body Systems}
\author{Stefano Longhi}
\email{stefano.longhi@polimi.it}
\affiliation{Dipartimento di Fisica, Politecnico di Milano, Piazza Leonardo da Vinci 32, I-20133 Milano, Italy}
\affiliation{IFISC (UIB-CSIC), Instituto de Fisica Interdisciplinar y Sistemas Complejos, E-07122 Palma de Mallorca, Spain}

\begin{abstract}
In non-Hermitian (NH) fermionic chains with spatially random imaginary gauge fields, single-particle eigenstates exhibit erratic NH skin localization with strongly fluctuating, subexponential decay, distinct from both conventional skin accumulation and Anderson localization. At finite filling, Pauli exclusion suppresses macroscopic accumulation and instead leads to fragmentation of the particle density, in sharp contrast to Hermitian disordered systems. Using exact free-fermion methods, we show that the density organizes into disconnected regions of high and low occupancy, forming spatially separated Fermi domains {\color{black} whose number $N_{\mathrm{FS}}$ scales subextensively with system size $L$}. Remarkably, fragmentation is not only an equilibrium property but also emerges dynamically in the NH evolution, unlike in Hermitian systems where such structures are not dynamically selected. This behavior admits a Coulomb-gas description in a stochastic potential, establishing a direct link between erratic NH skin localization and real-space fragmentation of many-body states.
\end{abstract}

\maketitle
\section{Introduction}
Non-Hermitian (NH) systems \cite{W1} exhibit phenomena and topological properties with no Hermitian counterpart \cite{W2,W3,W4,W5,W6,W7,W8,W9,W10,W11}, most notably the non-Hermitian skin effect (NHSE) \cite{N1}, whereby eigenstates accumulate at system boundaries under open boundary conditions \cite{N1,N2,N3,N4,N5,N6,N7,N8,N9,N10,N11,N12,N13,N14,N15,N16,N17,N18,N19,N20,N21,N22,N23,N24,N25,N26,N27,N27b,N28,N29}. The many-body consequences of non-Hermiticity have recently attracted significant attention, revealing a variety of unconventional behaviors in both interacting and noninteracting systems \cite{M1,M2,M3,M4,M5,M6,M6b,M7,M8,M9,M10,M11,M12,M13,M14,M15,M16,M17,M18,M18b,M19,M20,M21,M22,M22b,M23,M24,M25,M25b,M26,M27}. In fermionic systems, Pauli exclusion acts as an effective quantum degeneracy pressure that suppresses macroscopic boundary accumulation \cite{M2}, leading instead to an inhomogeneous density profile at finite filling \cite{M2,M4,M5}. In disorder-free models, this profile is characterized by a sharp crossover between regions of high and low occupancy, which can be interpreted as a real-space Fermi surface \cite{M2}. A natural question is how this structure responds to spatial disorder in the NH field. In the single-particle regime, stochastic asymmetric hopping that preserves global reciprocity gives rise to erratic non-Hermitian skin localization (ENHSL) \cite{E1,E2,E3,E4,E5,E6,E7}, characterized by strongly fluctuating localization patterns and sub-exponential localization governed by the extreme-value statistics of random walks \cite{E1}. However, the corresponding many-body behavior remains largely unexplored. In Hermitian disordered systems, Anderson localization produces spatially irregular density profiles for noninteracting fermions without fragmentation. It is therefore natural to ask whether disorder in the NH field leads to a fundamentally different behavior in the many-body regime.

Here we show that stochastic non-Hermiticity, giving rise to ENHSL at the single-particle level, fundamentally reshapes the many-body state compared to the Anderson-localized case. Using exact free-fermion methods, we find that the density does not merely exhibit spatial irregularity but instead fragments into multiple disconnected regions of high and low occupancy, corresponding to spatially separated Fermi domains. As the NH strength increases, the density profile evolves from weak fluctuations to sharply defined domain boundaries. A central result is that, at large NH strength, the number of such boundaries exhibits a subextensive scaling with system size $L$, {\color{black} reflecting the large-scale first-passage and record statistics of the
underlying random-walk gauge field.} This behavior admits a natural interpretation in terms of an effective Coulomb gas in a random potential, linking ENHSL to disorder-driven fragmentation of many-body states in real space. Our results reveal a new form of spatial organization in NH fermionic systems, where disorder and quantum degeneracy pressure combine to produce fragmented density profiles, extending the notion of the skin effect to many-body settings.

\section{Model and many-body ground state}
We consider a one-dimensional lattice of spinless fermions with asymmetric hopping [Fig.~1(a)], described by a NH Hatano-Nelson Hamiltonian \cite{M1,M2,M3,M4,M5,M6b,M10,M15}
\begin{equation}
H = J \sum_{n=1}^{L-1} \left( e^{g_n} c^\dagger_{n} c_{n+1} + e^{-g_n} c^\dagger_{n+1} c_n \right)
+ U \sum_{n=1}^{L-1} \hat{n}_n \hat{n}_{n+1},
\end{equation}
where $c_n^\dagger$ ($c_n$) creates (annihilates) a fermion at site $n$, $\hat{n}_n = c_n^\dagger c_n$, $g_n$ defines a spatially varying imaginary gauge field, and $U>0$ is the nearest-neighbor interaction energy. Such nonreciprocal hopping can be realized in ultracold fermionic atoms in optical lattices through engineered atom loss and dissipation channels \cite{M1}. 
The effective NH Hamiltonian (1) with inhomogeneous gauge field $g_n$ can be derived from a Lindblad master equation with  linear and non-local dissipators \cite{E2,L1,L2,L3,L4}, and describes the dynamics under conditional, no-jump evolution \cite{M1,M15,post1}. The limit of large $g_n$ and small $J$, with $J e^{|g_n|}$ finite, corresponds to a nearly unidirectional hopping. 
% The non-Hermitian Hamiltonian describes the quantum dynamics conditioned on the absence of quantum jumps within a quantum trajectory framework. 
We focus on open boundary conditions (OBC), under which non-Hermitian effects are most pronounced, although for globally reciprocal models equivalent results are obtained under periodic boundary conditions (PBC) \cite{E1}. Here, we primarily focus on the noninteracting case $U=0$, which allows for an exact treatment in terms of free fermions. The effects of nearest-neighbor interactions ($U>0$) are briefly discussed in \cite{Suppl}. % where it is shown that the particle density profile is not much affected by weak interaction strengths.
For uniform $g_n = g$, the model exhibits the NHSE at the single-particle level. In contrast, when $g_n$ is spatially random with zero mean, the system realizes stochastic asymmetric hopping with global reciprocity leading to ENHSL in the single-particle regime \cite{E1}. 
In the Hermitian limit ($g_n = 0$), the eigenstates under OBC are standing waves, $\phi_\alpha(n) = \sin\left(\frac{\pi \alpha n}{L+1}\right)$,
\begin{figure}[htbp]
	\centering
	\includegraphics[width=\linewidth]{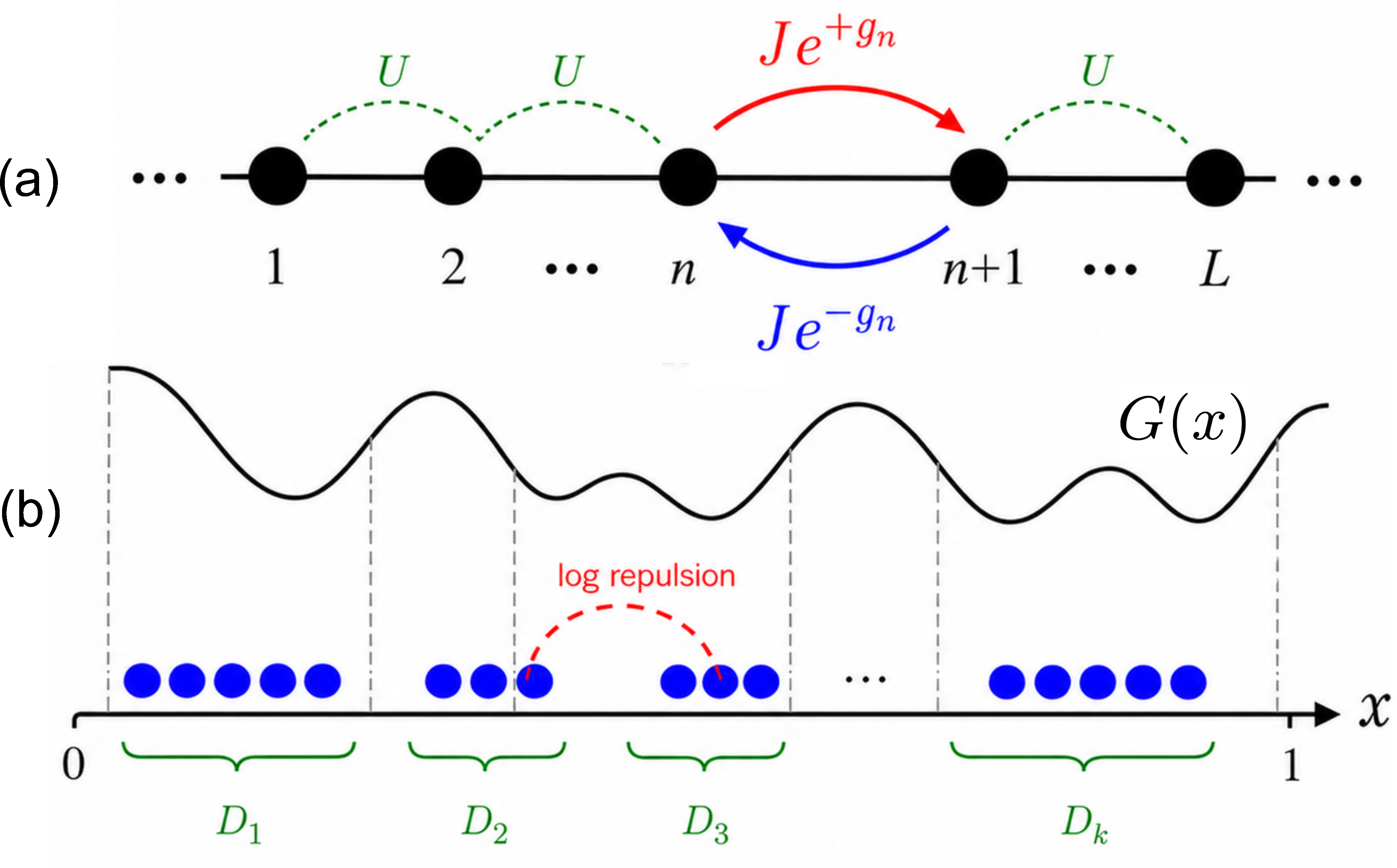}
	\caption{
Schematic illustration of the lattice model and its Coulomb-gas mapping. 
(a) One-dimensional fermionic lattice with asymmetric nearest-neighbor hopping and interaction. The left/right hopping amplitudes between sites $n$ and $n+1$ are $J e^{\pm g_n}$, where $\{g_n\}$ is the (generally stochastic) non-Hermitian gauge field. A nearest-neighbor interaction $U$ acts between adjacent sites (green dashed arcs). 
(b) Coulomb-gas representation in the continuum coordinate $x = n/L \in [0,1]$. Particles (blue dots) experience logarithmic repulsion and are
subject to the stochastic external potential $G(x)$, given by the
cumulative non-Hermitian field $G(x)=\sum_{y<x}g_y$. The equilibrium density forms $k$ disconnected regions $D_1, D_2, \dots, D_k$, corresponding to fragmented real-space Fermi domains.
}
\label{fig1}
\end{figure}
with eigenvalues $E_\alpha = 2J \cos\left(\frac{\pi \alpha}{L+1}\right)$, where $\alpha = 1,\dots,L$.
In the presence of non-Hermiticity, the energy spectrum remains invariant while the right eigenstates can be written as
\begin{equation}
\psi_\alpha(n) = \phi_\alpha(n) e^{-G(n)},
\end{equation}
where $G(n) = \sum_{m<n} g_m$ encodes the cumulative imaginary gauge field.
The many-body $N$-particle energies remain real and unaffected by the imaginary gauge fields, while the interplay between non-Hermiticity and many-body physics is entirely encoded in the structure of the eigenstates.
Observables are computed from the right-eigenstate Slater determinant, consistently with the definition of the many-body state \cite{M2,M4,M15}.
At filling $\nu = N/L$, we consider the many-body ground state constructed as a Slater determinant of the $N$ lowest-energy single-particle states,
\begin{equation}
\Psi(n_1,\dots,n_N) = \frac{1}{\sqrt{N!}} \det \left[ \psi_{\alpha_j}(n_i) \right],
\end{equation}
with $\alpha_j = L-j+1$. While we focus on the ground state for definiteness, the formalism readily extends to excited many-body states obtained by occupying different sets of single-particle orbitals, leading to qualitatively similar behavior. Using the factorized form of the single-particle orbitals, the (unnormalized) many-body wavefunction can be written as
\begin{equation}
\Psi(n_1,\dots,n_N) = e^{-\sum_{i=1}^N G(n_i)} \tilde{\Psi}(n_1,\dots,n_N),
\end{equation}
where $\tilde{\Psi}$ is the Slater determinant of the Hermitian problem. This representation highlights the competition between Pauli exclusion, i.e. quantum degeneracy pressure encoded in $\tilde{\Psi}$, and the spatially varying NH field.
 The local particle density is given by
\begin{equation}
\rho(n) = \frac{\langle \Psi | c_n^\dagger c_n | \Psi \rangle}{\langle \Psi | \Psi \rangle}.
\end{equation}
Denoting by $\Psi_{n\alpha} = \psi_\alpha(n)$ the matrix of occupied orbitals, the density can be written as
\begin{equation}
\rho(n) = (\Psi S^{-1} \Psi^\dagger)_{nn} = \sum_{\alpha=1}^{N} |Q_{n\alpha}|^2,
\end{equation}
where $S = \Psi^\dagger \Psi$ accounts for the non-orthogonality of the orbitals and $\Psi = Q R$ is the QR decomposition of $\Psi$ (see Appendix~A).
%The latter formulation is numerically stable, while remaining equivalent to the normalized many-body expectation value.

\vspace{0.2cm}

\begin{figure}[htbp]
	\centering
	\includegraphics[width=\linewidth]{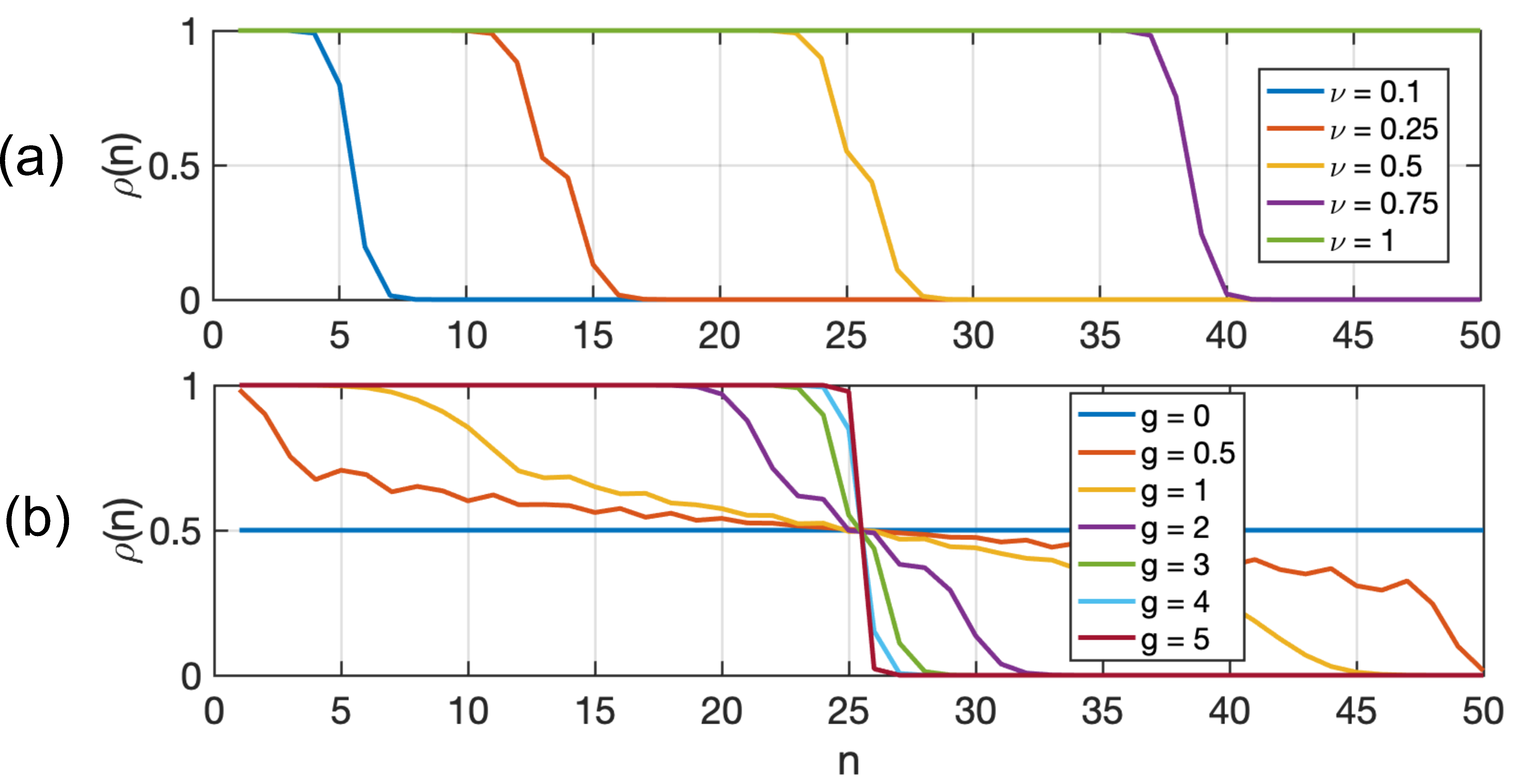}
	\caption{
Density profiles $\rho(n)$ for a one-dimensional lattice with $L=50$ sites, OBC, uniform non-Hermitian field $g_n=g$, and no interactions ($U=0$). 
(a) $\rho(n)$ for fixed $g=3$ and increasing filling $\nu$ (see legend), showing the progressive occupation of spatial domains. 
(b) $\rho(n)$ at half filling $\nu=1/2$ for increasing values of $g$ (see legend), illustrating the emergence of a sharp real-space Fermi surface as $g$ is increased.
}\label{fig2}
\end{figure}

\section{Particle density fragmentation}
We begin by recalling the structure of the many-body ground state for a
uniform imaginary gauge field $g_n=g$ \cite{M2}. Under OBC, Pauli exclusion
prevents macroscopic particle accumulation associated with the NHSE and
instead produces an inhomogeneous density profile that deviates from
exponential skin accumulation, approaching a constant density as
$\nu\to1$ [see Fig.~2(a)]. The many-body signature of the NHSE is embodied
in a non-vanishing value of the dipole moment
$P=\sum_{n=1}^L n[\rho(n)-\nu]$ \cite{M15}.

For sufficiently large $g$ and finite filling $0<\nu<1$ (away from the
dilute and fully filled limits), the density develops a sharp spatial
crossover between regions of high and low occupancy [Fig.~2(b)], which can
be interpreted as a real-space Fermi surface \cite{M2}.
\textcolor{black}{Its emergence can be understood as the competition between
the NHSE, which tends to accumulate all single-particle states toward the
same boundary, and Pauli exclusion, or equivalently quantum degeneracy
pressure, which prevents the fermions from occupying the same states and
forces them to spread progressively into the bulk \cite{M2}.}
The position of this boundary is set by the filling,
$x_F\simeq\nu$ [Fig.~2(a)].
\textcolor{black}{Consequently, increasing $\nu$ primarily displaces the
real-space Fermi boundary from the accumulation edge toward the opposite
edge. In the strong-skin regime, where the densities on the two sides
approach $\rho\simeq1$ and $\rho\simeq0$, respectively, the profiles at
different fillings are therefore approximately related by a spatial
translation of the domain wall, as visible in Fig.~2(a). This correspondence
is not exact, since the shape and width of the interface retain some
dependence on the occupied single-particle orbitals, particularly at moderate
$g$.}

We now investigate how this structure is modified in the presence of spatially
random asymmetric hopping with zero mean. Specifically, we consider $g_n$ as a
stochastic variable drawn from a Bernoulli distribution, taking values $\pm g$
with equal probability.
\textcolor{black}{Figure~3 illustrates the resulting density fragmentation from
two complementary perspectives: panel (a) shows its evolution with NH strength
$g$ at half filling, whereas panel (b) shows its dependence on the filling
fraction $\nu$ at fixed $g=5$. In both panels the profiles are vertically
displaced for clarity.}
The disappearance of the NHSE is indicated by the vanishing of the dipole
moment $P$. \textcolor{black}{At half filling and weak $g$,} the density
$\rho(n)$ exhibits small, irregular fluctuations around its average value
$1/2$, reflecting the interplay of non-Hermiticity and interference
\textcolor{black}{[Fig.~3(a)]}. This behavior is qualitatively similar to that
found in Hermitian disordered systems, where Anderson localization leads to
spatially irregular but smooth density profiles \cite{Suppl}. As $g$ is
increased, however, these fluctuations become progressively sharper, and the
density reorganizes into well-defined regions of high and low occupancy,
separated by sharp interfaces that can be viewed as real-space Fermi surfaces
\textcolor{black}{[Fig.~3(a)]}. This emergence of sharply defined domains is
absent in the Anderson-localized case and signals a qualitatively distinct
many-body regime driven by non-Hermiticity.
\textcolor{black}{The fragmented structure is not restricted to half filling.
Figure~3(b) shows the density for the same realization of the stochastic
imaginary-gauge field at fixed $g=5$ and increasing filling fraction $\nu$.
As particles are added, the domains evolve within the same underlying
stochastic landscape: favorable regions are progressively populated and
expanded, while neighboring occupied domains can eventually merge.
Fragmentation therefore persists over a broad range of fillings, with $\nu$
controlling the spatial arrangement of the real-space Fermi domains rather
than simply enhancing fragmentation monotonically. In this sense, the
filling-dependent displacement of the single real-space Fermi boundary in
the uniform Hatano--Nelson chain [Fig.~2(a)] is generalized, in the stochastic
system, to a filling-dependent reorganization of multiple real-space Fermi
domains. In the following, we restrict the analysis to half filling,
$\nu=1/2$.}

Operationally, we identify these interfaces through
\textcolor{black}{crossings of the density threshold $\rho_{\rm th}=1/2$
between neighboring lattice sites, i.e.,
\begin{equation}
[\rho(n)-1/2][\rho(n+1)-1/2]<0,
\end{equation}}
and define $N_{\mathrm{FS}}$ as the number of such
\textcolor{black}{interfaces}.
A key quantity is the scaling of $N_{\mathrm{FS}}$ with system size $L$.
\textcolor{black}{Figure~4(a) shows the disorder-averaged
$\langle N_{\mathrm{FS}}\rangle$ versus $L$ for representative values of the
gauge strength $g$, with each data point averaged over $10^3$ realizations of
the stochastic field $g_n$. The lines proportional to $L$ and $L^{1/2}$ are
shown only as reference scalings.}
For small $g$, the density $\rho(n)$ exhibits weak, random fluctuations around
its average value (e.g. $1/2$ at half filling), leading to an apparent
extensive scaling $N_{\mathrm{FS}}\sim L$ \textcolor{black}{, as expected for
microscopic disorder-induced density fluctuations}. As $g$ is increased, the
density reorganizes into multiple disconnected regions of high and low
occupancy, \textcolor{black}{and a pronounced departure from extensivity
develops.}
\textcolor{black}{To quantify this crossover, we introduce an effective
finite-size exponent $\gamma_{\rm eff}$ through
\begin{equation}
N_{\mathrm{FS}}\sim L^{\gamma_{\rm eff}},
\end{equation}
extracted over the common large-size window $L\geq100$. Figure~4(b) shows
the resulting $\gamma_{\rm eff}$ versus $g$. It evolves from values close to
unity at weak $g$ toward a broad strong-NH regime with
$\gamma_{\rm eff}\simeq0.45$--$0.5$. The residual variation in this regime
reflects finite-size corrections, and the numerical data should therefore not
be interpreted as establishing an exact asymptotic exponent of $1/2$.
Rather, the robust numerical result is the crossover from extensive to
strongly subextensive scaling. This smooth evolution does not indicate a
sharp phase transition at a finite critical value of $g$.}
\textcolor{black}{Fragmentation is characterized by a number of
interfaces that diverges while remaining subextensive,
\begin{equation}
N_{\mathrm{FS}}\rightarrow\infty,
\qquad
\frac{N_{\mathrm{FS}}}{L}\rightarrow0
\qquad (L\rightarrow\infty).
\end{equation}
Consequently, the characteristic separation between interfaces,
$\ell_{\rm dom}\sim L/N_{\mathrm{FS}}$, also diverges in the thermodynamic
limit. The strongly subextensive behavior observed in Fig.~4 therefore
provides the numerical signature of {\em density fragmentation}. As discussed
below, the square-root scale suggested by the strong-NH data has a natural
interpretation in terms of the first-passage and record statistics of the
underlying random-walk gauge landscape.}
We further characterize the spatial structure of $\rho(n)$ by considering the
participation ratio
$\mathrm{PR}_\rho=(\sum_n\rho(n))^2/\sum_n\rho(n)^2$, which measures the
effective number of sites occupied by the density profile. Together with the
scaling behavior of $N_{\rm FS}$, it confirms the fragmented nature of the
density, which is further reflected in the emergence of spatially structured
correlations \cite{Suppl}.

\textcolor{black}{Although the above analysis exploits the free-fermion
structure at $U=0$, density fragmentation is intrinsically a finite-density
fermionic effect rather than a single-particle phenomenon. The stochastic
imaginary-gauge field tends to concentrate the single-particle orbitals in
favorable spatial regions, while Pauli exclusion prevents their macroscopic
occupation by many fermions, producing the spatially separated Fermi domains
described above. Genuine particle-particle interactions are therefore not
required for fragmentation, but can modify this competition.
Importantly, the imaginary-gauge transformation itself remains exact in the
presence of the nearest-neighbor density-density interaction considered here.
Indeed, under OBC one can introduce the many-body similarity transformation
\begin{equation}
\hat S=\exp\left[-\sum_n G(n)\hat n_n\right],
\end{equation}
which gauges away the asymmetric hopping. Since $\hat S$ is a function only
of the mutually commuting local occupation operators,
$[\hat S,\hat n_n]=0$, and therefore
\begin{equation}
\hat S\,\hat n_n\hat n_{n+1}\,\hat S^{-1}
=\hat n_n\hat n_{n+1}.
\end{equation}
Hence, even for spatially varying $G(n)$, the interacting NH Hamiltonian is
similar under OBC to the corresponding Hermitian interacting fermionic chain.
Equivalently, its right many-body eigenstates retain the common
imaginary-gauge factor,
\begin{equation}
\Psi_U(n_1,\ldots,n_N)
=
\exp\left[-\sum_{j=1}^{N}G(n_j)\right]
\widetilde{\Psi}_U(n_1,\ldots,n_N),
\end{equation}
where $\widetilde{\Psi}_U$ is the corresponding eigenstate of the Hermitian
interacting problem. Thus, interactions do not remove the stochastic gauge
reweighting underlying fragmentation, but modify the correlated Hermitian
many-body state on which this reweighting acts.
To assess the robustness of the resulting fragmented structure, we have
performed exact-diagonalization calculations including the nearest-neighbor
interaction $U$ \cite{Suppl}. For the accessible system sizes,
weak-to-moderate repulsive interactions modify the detailed density profile
but leave the fragmented high- and low-density domains clearly visible,
showing that fragmentation is not singular to the exactly noninteracting
limit.
At sufficiently strong repulsion, a qualitatively different regime may
emerge. In the uniformly nonreciprocal Hatano--Nelson chain, nearest-neighbor
repulsion is known to compete with the NH-induced real-space Fermi surface by
favoring charge-density-wave (CDW) correlations \cite{M2}. A similar
competition can be anticipated here: at half filling, strong repulsion
penalizes consecutive particle occupation and may generate short-range CDW
ordering within the particle-rich regions selected by the stochastic gauge
landscape. This suggests the possibility of a richer regime in which the
random imaginary-gauge field controls the large-scale fragmented envelope,
while interactions produce shorter-scale density ordering within individual
domains. Determining how this competition affects the thermodynamic scaling
of fragmentation constitutes an interesting direction for future
investigation.}

\begin{figure*}
	\centering
	\includegraphics[width=0.9\linewidth]{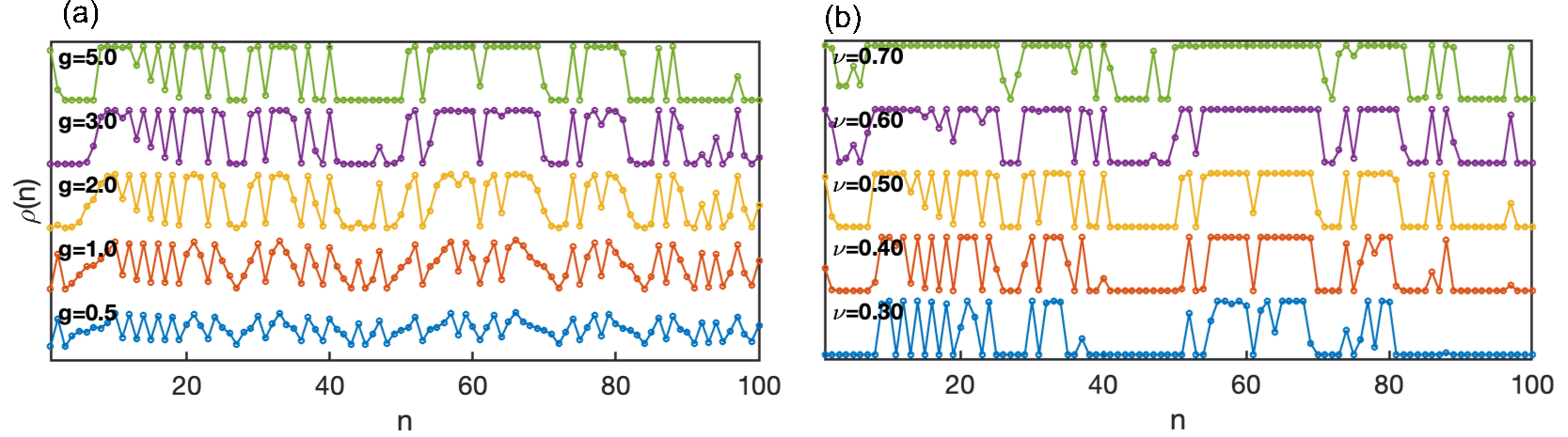}
	\caption{
{\color{black} Density fragmentation in the stochastic non-Hermitian chain.
(a) Density profiles $\rho(n)$ at half filling for increasing non-Hermitian
strength $g$ and $U=0$. The profiles are vertically
displaced for clarity. Increasing $g$
progressively sharpens the fragmented high- and low-density domains.
(b) Density profiles for a fixed realization of the stochastic imaginary
gauge field at $g=5$, $U=0$ and filling fractions
$\nu=0.30,\,0.40,\,0.50,\,0.60$, and $0.70$. Fragmentation persists over a broad range of fillings,
while increasing $\nu$ progressively populates, expands, and merges the
domains selected by the same stochastic landscape.}}
\label{fig3}
\end{figure*}
\begin{figure}
	\centering
	\includegraphics[width=\linewidth]{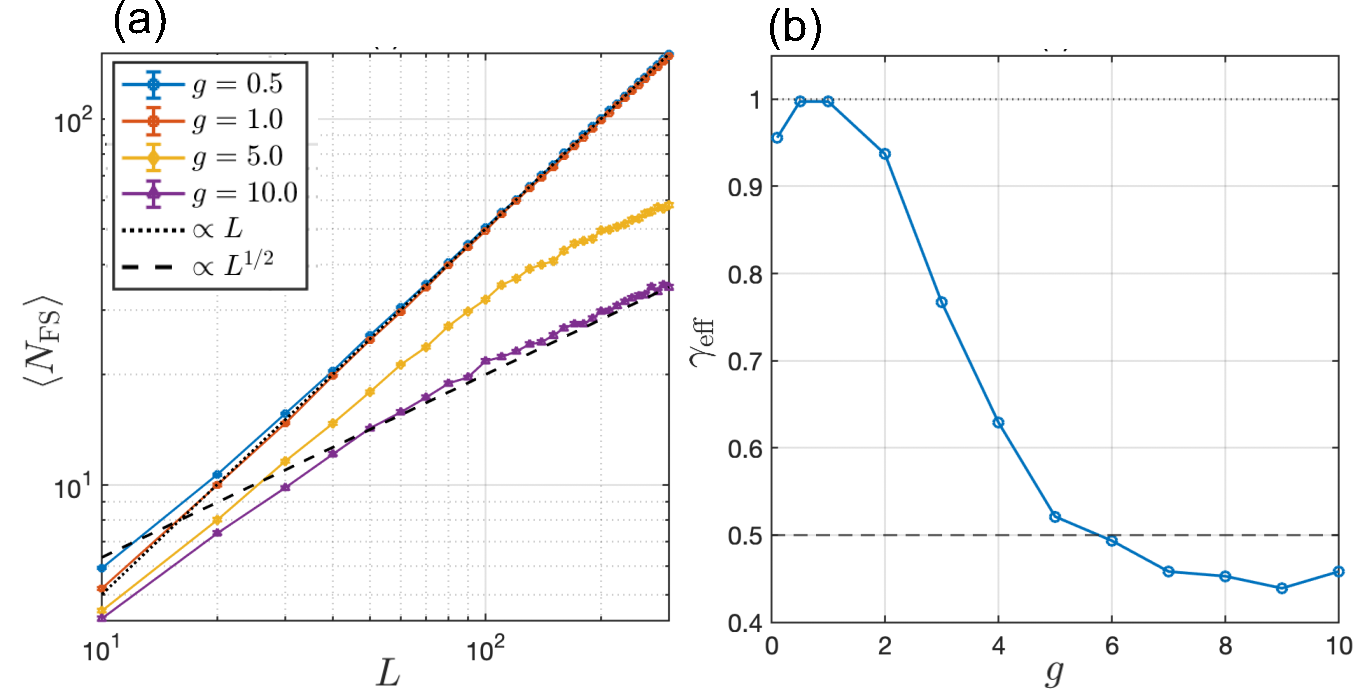}
	\caption{
Scaling of the number of real-space Fermi interfaces in the stochastic
non-Hermitian chain at half filling. \color{black}{
(a) Disorder-averaged number of density interfaces
$\langle N_{\rm FS}\rangle$ versus system size $L$ for representative values
of the imaginary-gauge strength $g$. Each point is averaged over $10^3$
independent realizations of the Bernoulli stochastic field.
\textcolor{black}{Interfaces are identified directly from crossings of the
density threshold $\rho_{\rm th}=1/2$ between neighboring sites. The dotted
and dashed black lines indicate the reference scalings
$N_{\rm FS}\propto L$ and $N_{\rm FS}\propto L^{1/2}$, respectively, and
are shown as guides to the eye.}
(b) \textcolor{black}{Effective finite-size exponent $\gamma_{\rm eff}$,
defined by $N_{\rm FS}\sim L^{\gamma_{\rm eff}}$ and extracted over the
common large-size window $L\geq100$, as a function of $g$. The horizontal
dotted and dashed lines indicate $\gamma_{\rm eff}=1$ and
$\gamma_{\rm eff}=1/2$, respectively. Increasing $g$ produces a crossover
from approximately extensive scaling at weak non-Hermiticity to a strongly
subextensive regime at large $g$.}
} }
\label{fig4}	\label{fig4}
\end{figure}

\section{Coulomb-gas description}
\noindent
In the continuum limit, the many-body problem admits a natural description as
a Coulomb gas in a stochastic potential $G(x)$, where Pauli exclusion induces
an effective logarithmic repulsion. Within this framework, the density can
develop disconnected support, providing a continuum counterpart of the lattice
fragmentation.
This mapping follows from the structure of the many-body wavefunction.
Although typically derived for continuum fermions \cite{C0,C0bis}, here it
emerges from the polynomial form of lattice eigenstates under open boundary
conditions and becomes exact at large $L$ (see Appendix~B). Introducing
continuum coordinates $x_i=n_i/L\in(0,1)$, the squared amplitude of the Slater
determinant reads approximately
\begin{equation}
|\Psi(x_1,\dots,x_N)|^2 \propto
\exp\left(-2 \sum_i G(x_i)\right)
\prod_{i<j}(x_i-x_j)^2,
\end{equation}
where $G(x)$ is the cumulative NH field. This defines a classical Gibbs weight
$P\propto e^{-\mathcal{H}}$ with effective Hamiltonian
\begin{equation}
\mathcal{H}=
-2\sum_{i<j}\log|x_i-x_j|
+2\sum_i G(x_i),
\end{equation}
corresponding to a one-dimensional Coulomb gas in an external potential
\cite{C1,C2,C3,C4}; see Fig.~1(b).

The coarse-grained density $\rho(x)$ is the equilibrium configuration of this
gas. For smooth $G(x)$, it has a single connected support, corresponding to
one real-space Fermi surface. In contrast, a stochastic $G(x)$ develops
multiple extrema, leading to fragmentation into disconnected intervals
[Fig.~1(b)].
In the large-$g$ regime, $G(x)$ exhibits deep minima separated by barriers.
\textcolor{black}{Unlike a smooth confining potential, however, a random-walk
landscape is non-smooth and its relevant valleys are not generically
quadratic. The competition between localization near these valleys and the
logarithmic fermionic repulsion produces weakly overlapping density droplets
separated by depleted regions, corresponding in real space to alternating
occupied and empty domains.}
\textcolor{black}{Importantly, the number of ordinary local extrema of a
random-walk landscape is extensive, $N_{\rm ext}=O(L)$, and therefore cannot
itself account for fragmentation. The Coulomb-gas/Pauli competition
coarse-grains these microscopic fluctuations: shallow neighboring extrema do
not generally produce independent density domains, and the relevant
structures are instead the sparse hierarchy of deep, record-like excursions
of $G(n)$. This distinction can be formalized, in the Brownian limit, through
the $h$-extrema construction of Neveu and Pitman \cite{NeveuPitman1989},
which selects extrema surviving a prescribed excursion depth and thereby
filters out the abundance of microscopic extrema.}
\textcolor{black}{A complementary characterization is provided by record
statistics. A lower record occurs whenever
\begin{equation}
G(n)<\min_{m<n}G(m).
\end{equation}
For a symmetric one-dimensional random walk, the probability of establishing
a new record at step $n$ decays asymptotically as $n^{-1/2}$. Consequently,
the mean number of record events accumulated over a chain of length $L$
scales as \cite{MajumdarZiff2008}
\begin{equation}
\langle N_{\rm rec}(L)\rangle\sim L^{1/2}.
\end{equation}
Thus, $L^{1/2}$ provides a natural subextensive statistical scale for the
large-scale valleys and excursions of the stochastic gauge landscape.}
\textcolor{black}{This argument should not be interpreted as an exact
one-to-one identification of record events with density domains: several
successive records may belong to the same valley, and whether a given valley
supports a distinct domain also depends on filling and on the logarithmic
repulsion. Accordingly, the Coulomb-gas mapping does not by itself constitute
an exact derivation of $N_{\rm FS}\propto L^{1/2}$. Rather, it explains how
the extensive microscopic structure of $G(n)$ is coarse-grained by the
fermionic many-body state, while universal first-passage and record statistics
identify the natural square-root scale of the remaining large-scale
fluctuations. Consistently with the finite-size analysis of Fig.~4, the robust
signature of fragmentation is the strongly subextensive growth of
$N_{\rm FS}$, while the square-root law provides the theoretically motivated
asymptotic scale of the stochastic landscape.}

\vspace{0.2cm}

\begin{figure}[htbp]
	\centering
	\includegraphics[width=\linewidth]{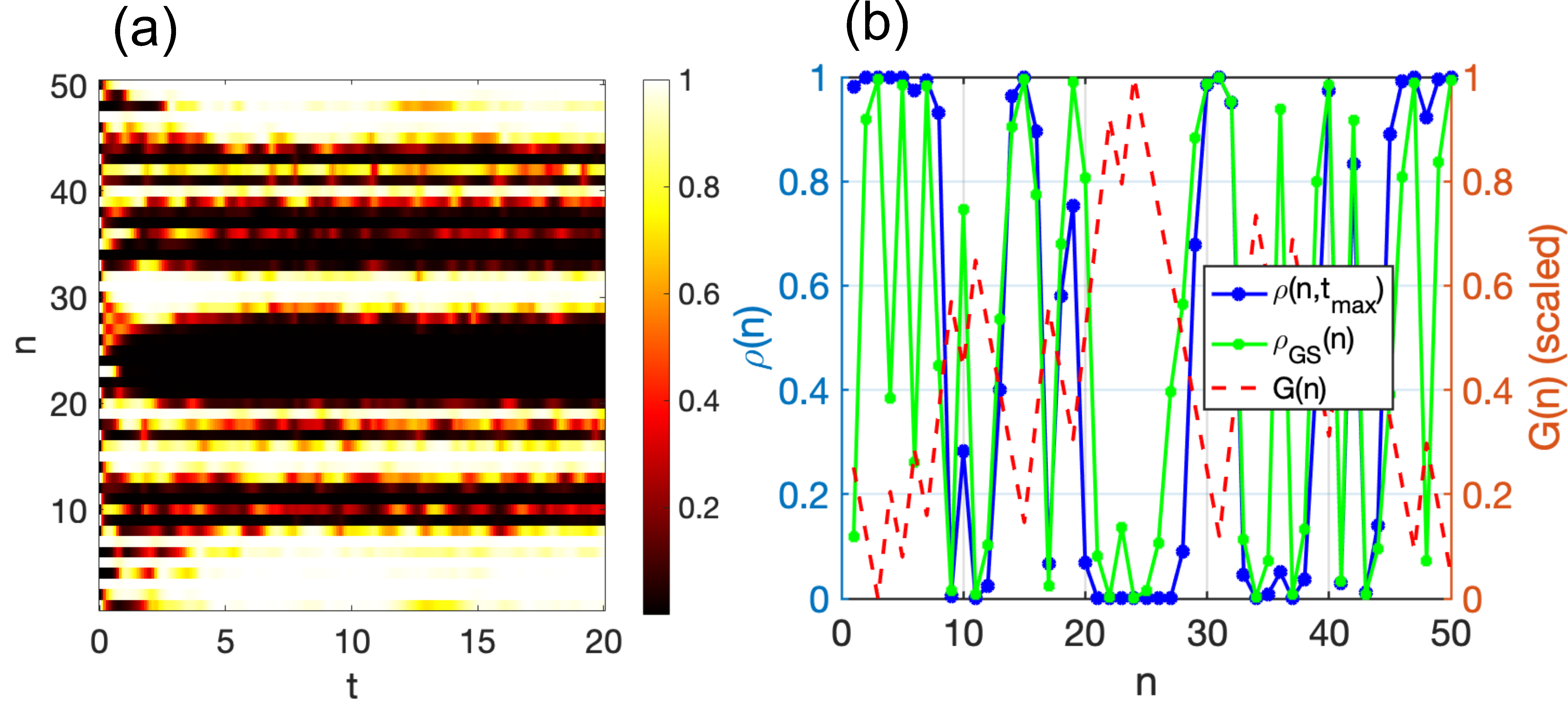}
	\caption{(a) Spatiotemporal evolution of the density $\rho(n,t)$ for {\color{black} $L=50$} at half filling, starting from a charge-density-wave state. Dynamics is governed by the stochastic NH Hamiltonian (1) with $g=2$, $J=1$, $U=0$ and final time $t_{\mathrm{max}}=20$. (b) Density at $t_{\mathrm{max}}$ (blue circles) compared with the ground-state profile $\rho_{\mathrm{GS}}(n)$ (green squares) and the cumulative field $G(n)$ (red dashed line). The dynamics leads to fragmented domains closely matching the ground state, with occupied (depleted) regions aligned with minima (maxima) of $G(n)$.}
\label{fig5}
\end{figure}
\section{Dynamical emergence of fragmentation}

\textcolor{black}{The fragmented spatial organization found in the many-body
eigenstates can also emerge dynamically from an initially unfragmented state.}
We consider a lattice of size $L$ at half filling, initialized in a
charge-density-wave state with alternating site occupation. The many-body
wavefunction evolves as \cite{M1,M12,M15,M18}
\[
|\Psi(t)\rangle=
\frac{e^{-iHt}|\Psi(0)\rangle}
{\|e^{-iHt}|\Psi(0)\rangle\|},
\]
corresponding to no-jump quantum trajectory dynamics. For noninteracting
fermions, the evolution is obtained by propagating the occupied
single-particle orbitals and reconstructing the Slater determinant.
Numerically, we evolve the orbital matrix $\Psi(t)$ and compute the density
from the projector
\[
\rho(n,t)=(\Psi S^{-1}\Psi^\dagger)_{nn},
\qquad
S=\Psi^\dagger\Psi .
\]

Figure~5(a) shows the evolution of $\rho(n,t)$ for a representative
realization of the stochastic field $g_n$. The initial short-range
oscillations rapidly wash out, and the density
\textcolor{black}{develops a coarse-grained spatial organization}
with well-defined high- and low-density regions, signaling fragmentation.
Figure~5(b) compares the density at the final simulation time with the
ground-state profile and the cumulative field $G(n)$. The close
\textcolor{black}{spatial correspondence shows that the dynamically generated
density is organized by the same stochastic gauge landscape as the
fragmented eigenstates:} particles preferentially accumulate near minima of
$G(n)$ and depleted regions occur near its maxima, consistent with the
Coulomb-gas picture.
\textcolor{black}{This dynamical behavior should not be interpreted as
relaxation toward the ground state or toward a dynamical attractor. Under
OBC, the NH Hamiltonian is related to its Hermitian parent by a similarity
transformation and consequently has a real spectrum. There is therefore no
generic exponential spectral selection of a particular eigenstate during
the normalized no-jump evolution. Instead, all single-particle right
eigenstates share the same stochastic gauge factor,
$\psi_\alpha(n)=e^{-G(n)}\phi_\alpha(n)$, and the corresponding many-body
eigenstates inherit the common factor
\begin{equation}
\exp\left[-\sum_{j=1}^{N}G(n_j)\right].
\end{equation}
Although their detailed density profiles are not identical, they are thus
subject to the same stochastic spatial reweighting and share the same
underlying tendency toward fragmented spatial organization.}
\textcolor{black}{A generic time-dependent state remains a coherent
superposition of these eigenstates. Since their eigenenergies are real, the
relative components do not generically decay away, and temporal oscillations
may persist at long times. Figure~5 therefore demonstrates the
\emph{dynamical emergence of fragmentation}, rather than convergence to a
unique stationary fragmented state.}

While spatially inhomogeneous densities and domain formation are known in
various many-body Hermitian settings, such as interaction-driven phase
separation, inhomogeneous trapping potentials, and Anderson-localized
systems, the fragmentation observed here arises from a distinct mechanism
\textcolor{black}{associated with the common non-unitary spatial reweighting of
the eigenstates.} For example, in Hermitian systems with spatially varying
on-site potentials, although stationary many-body eigenstates may exhibit
\textcolor{black}{strongly structured or multi-domain densities}, the unitary
dynamics does not \textcolor{black}{dynamically generate the same persistent
coarse-grained fragmented organization}: the density remains oscillatory
\textcolor{black}{without developing the characteristic NH domain structure}
\cite{Suppl}.

Finally, the mechanism discussed here is not restricted to
the stochastic Hatano--Nelson chain. As detailed in the Supplemental
Material, it extends to the broad class of NH Hamiltonians obtained from
Hermitian parent systems through spatially varying imaginary-gauge
transformations, including models with long-range hopping and multiband
structure \cite{Suppl}.
\vspace{0.2cm}

\section{Conclusions}

Spatially random NH hopping reshapes many-body fermionic states, giving rise
to fragmentation of the particle density into disconnected regions. This
phenomenon originates from the interplay between Pauli exclusion and
stochastic non-Hermiticity, and provides a many-body manifestation of erratic
NH skin localization recently predicted and observed in the single-particle regime \cite{
E1,E5,E6,E7}. 
\textcolor{black}{Its defining signature is a diverging
but subextensive number of real-space Fermi interfaces,
$N_{\rm FS}\rightarrow\infty$ with $N_{\rm FS}/L\rightarrow0$, corresponding
to a characteristic domain size that also diverges with system size. In the
strong-NH regime the numerically extracted effective exponents are close to
the square-root scale, $L^{1/2}$, naturally suggested by the universal
first-passage and record statistics of the underlying random-walk gauge
landscape.}

Within a Coulomb-gas framework, fragmentation reflects the emergence of
spatially localized domains pinned by the stochastic potential.
\textcolor{black}{The logarithmic fermionic repulsion provides a natural
coarse-graining mechanism that filters out the extensive microscopic extrema
of the random walk and makes the density sensitive to its large-scale valleys
and excursions. Fragmentation is not restricted to the ground state or to
half filling, but emerges over a broad range of particle densities and can
also develop dynamically from an initially unfragmented state. Since the OBC
spectrum is real, this dynamical emergence should not be interpreted as
relaxation toward a unique attractor: coherent temporal oscillations can
persist while the density acquires the characteristic fragmented spatial
organization imposed by the stochastic gauge landscape.}

\textcolor{black}{Although the scalable analysis exploits the free-fermion
structure, fragmentation is intrinsically a finite-density fermionic effect:
explicit particle-particle interactions are not required, whereas Pauli
exclusion is essential. Exact-diagonalization results further show that the
fragmented density structure survives weak-to-moderate nearest-neighbor
repulsion. At stronger interactions, competition with interaction-induced
charge-density-wave correlations may lead to richer behavior, possibly
combining a large-scale fragmented envelope with shorter-scale density
ordering.}

\textcolor{black}{Finally, the mechanism is not specific to the stochastic
Hatano--Nelson chain, but extends to a broad class of globally reciprocal NH
lattices generated by spatially varying imaginary-gauge transformations of
Hermitian parent Hamiltonians. These results identify stochastic
imaginary-gauge fields and Fermi statistics as a general route to
subextensive spatial organization in NH many-body systems, and open
directions toward interacting, multiband, long-range, and higher-dimensional
settings.}

\begin{acknowledgments}
This work was supported by the Spanish Agencia Estatal de Investigaci\'on through grant CEX2021-001164-M (Mar\'ia de Maeztu Program for Units of Excellence in R\&D), funded by MICIU/AEI/10.13039/501100011033.
\end{acknowledgments}

\section*{Data availability}
There are no publicly available research data or software supporting this manuscript. Requests for further information or data should be sent to the author.

\appendix

\section{Slater determinant structure and one-body correlation matrix}
\noindent
We consider the many-body ground state constructed from the single-particle orbitals
\begin{equation}
\psi_\alpha(n) = \phi_\alpha(n) e^{-G(n)},
\end{equation}
where $\phi_\alpha(n)$ are the Hermitian eigenstates and $G(n)$ is the cumulative imaginary gauge field. $\psi_\alpha(n)$ can be regarded as the elements of a $L \times N$ matrix $ \Psi$. 
The $N$-particle wavefunction is given by the Slater determinant
\begin{equation}
\Psi(n_1,\dots,n_N) = \frac{1}{\sqrt{N!}} \det \left[ \psi_{\alpha_j}(n_i) \right].
\end{equation}
Using the multiplicative structure of $\psi_\alpha(n)$, one can factor out a site-dependent contribution from each row of the determinant, yielding
\begin{equation}
\Psi(n_1,\dots,n_N)
= \exp\left(-\sum_{i=1}^N G(n_i)\right)\,
\tilde{\Psi}(n_1,\dots,n_N),
\end{equation}
where $\tilde{\Psi}$ is the Slater determinant of the Hermitian problem.  
This factorization shows that the NH effects enter only through a collective coordinate depending on all particle positions.
\noindent
Introducing fermionic creation operators associated with the single-particle orbitals,
\begin{equation}
d_\alpha^\dagger = \sum_n \psi_\alpha(n)\, c_n^\dagger,
\end{equation}
the many-body state can be written as
\begin{equation}
|\Psi\rangle = \prod_{\alpha=1}^N d_\alpha^\dagger |0\rangle.
\end{equation}
In general, the orbitals $\psi_\alpha(n)$ are not orthonormal, and therefore the operators $d_\alpha$ satisfy
\begin{equation}
\{ d_\alpha, d_\beta^\dagger \} = S_{\alpha\beta},
\end{equation}
where $S = \Psi^\dagger \Psi$ is the overlap matrix with elements
\begin{equation}
S_{\alpha\beta} = \sum_n \psi_\alpha^*(n)\psi_\beta(n).
\end{equation}
Using the Slater determinant structure of the many-body state, the one-body correlation matrix
\begin{equation}
C_{nm} = \frac{\langle \Psi | c_n^\dagger c_m | \Psi \rangle}{\langle \Psi | \Psi \rangle}
\end{equation}
can be expressed directly in terms of the occupied orbitals. In particular, using the standard overlap formula for Slater determinants (L\"owdin's theorem \cite{uff}), which expresses many-body overlaps as determinants of single-particle overlaps, one finds for non-orthogonal orbitals
\begin{equation}
C_{nm} = \sum_{\alpha,\beta}
\psi_\alpha(n)\,(S^{-1})_{\alpha\beta}\,\psi_\beta^*(m),
\end{equation}
or, in matrix form,
\begin{equation}
C = \Psi S^{-1} \Psi^\dagger,
\end{equation}
where $\Psi_{n\alpha}=\psi_\alpha(n)$. This result can be equivalently derived from the fermionic anticommutation relations of the operators $d_\alpha$ \cite{uff2}.
The local density then follows as
\begin{equation}
\rho(n) = C_{nn}.
\end{equation}
For numerical purposes, it is advantageous to avoid the explicit inversion of the overlap matrix. Writing the QR decomposition
\begin{equation}
\Psi = Q R,
\end{equation}
where $Q^\dagger Q = I$ and $R$ is upper triangular, one has $S = R^\dagger R$, which implies
\begin{equation}
C = \Psi S^{-1} \Psi^\dagger = Q Q^\dagger.
\end{equation}
Therefore,
\begin{equation}
\rho(n) = \sum_{\alpha=1}^N |Q_{n\alpha}|^2.
\end{equation}
This formulation is numerically stable and explicitly enforces the Pauli constraint $0 \le \rho(n) \le 1$, while remaining equivalent to the normalized many-body expectation value.

%Using the Slater determinant structure of the many-body state, the one-body correlation matrix
%\begin{equation}
%C_{nm} = \frac{\langle \Psi | c_n^\dagger c_m | \Psi \rangle}{\langle \Psi | \Psi \rangle}
%\end{equation}
%can be expressed directly in terms of the occupied orbitals. In particular, invoking the standard overlap formula for Slater determinants (L\"owdin's theorem \cite{uff}), which expresses many-body overlaps as determinants of single-particle overlaps -- and admits an equivalent derivation in operator form based on fermionic anticommutation relations \cite{uff2} -- one finds for non-orthogonal orbitals
%\begin{equation}
%C_{nm} = \sum_{\alpha,\beta}
%\psi_\alpha(n)\,(S^{-1})_{\alpha\beta}\,\psi_\beta^*(m),
%\end{equation}
%or, in matrix form,
%\begin{equation}
%C = \Psi S^{-1} \Psi^\dagger,
%\end{equation}
%where $\Psi_{n\alpha}=\psi_\alpha(n)$. The local density then follows as
%\begin{equation}
%\rho(n) = C_{nn}.
%\end{equation}
%For numerical purposes, it is advantageous to avoid the explicit inversion of the overlap matrix. Writing the QR decomposition
%\begin{equation}
%\Psi = Q R,
%\end{equation}
%where $Q^\dagger Q = I$ and $R$ is upper triangular, one has $S = R^\dagger R$, which implies
%\begin{equation}
%C = \Psi S^{-1} \Psi^\dagger = Q Q^\dagger.
%\end{equation}
%Therefore,
%\begin{equation}
%\rho(n) = \sum_{\alpha=1}^N |Q_{n\alpha}|^2.
%\end{equation}
%This formulation is numerically stable and explicitly enforces the Pauli constraint $0 \le \rho(n) \le 1$, while remaining equivalent to the normalized many-body expectation value.

\section{Continuum limit and mapping to a Coulomb gas}
\noindent
The many-body Hermitian wavefunction constructed from the standing-wave eigenfunctions,
\begin{equation}
\phi_\alpha(n) = \sqrt{\frac{2}{L+1}} \sin\left(\frac{\pi \alpha n}{L+1}\right),
\end{equation}
reads
\begin{equation}
\tilde{\Psi}(n_1,\dots,n_N)
= \det\left[\phi_{\alpha_j}(n_i)\right].
\end{equation}
This determinant can be rewritten in a form reminiscent of a Vandermonde structure by exploiting the identity
\begin{equation}
\sin[(m+1)\theta] = \sin\theta \, U_m(\cos\theta),
\end{equation}
where $U_m(x)$ is the Chebyshev polynomial of the second kind. Setting $\theta_{n_i}=\pi n_i/(L+1)$ and $\alpha_j=m_j+1$, one obtains
\begin{equation}
\tilde{\Psi}(n_1,\dots,n_N)
\propto
\prod_{i=1}^N \sin\theta_{n_i}
\det\left[ U_{m_j}(\cos\theta_{n_i}) \right].
\end{equation}
Since $U_m(x)$ is a polynomial of degree $m$, the determinant $\det[U_{m_j}(x_i)]$ is an \emph{alternant}, and therefore proportional to a Vandermonde determinant in the variables $x_i=\cos\theta_{n_i}$. This yields
\begin{equation}
\tilde{\Psi}(n_1,\dots,n_N)
\propto
\prod_{i<j}
\left[
\cos\theta_{n_i} - \cos\theta_{n_j}
\right]
\prod_{i=1}^N \sin\theta_{n_i},
\end{equation}
where $\theta_n = \pi n/(L+1)$.
This representation makes explicit the antisymmetric structure of the wavefunction through pairwise differences of $\cos\theta_{n_i}$.
To connect with a continuum description, we introduce scaled coordinates $x_i = n_i/L \in (0,1)$. In the large-$L$ limit,
\begin{equation}
\cos\theta_{n_i} - \cos\theta_{n_j}
\simeq
-\sin(\bar{\theta})\,(\theta_{n_i} - \theta_{n_j})
\propto (x_i - x_j),
\end{equation}
where $\bar{\theta}$ is an average angle between $\theta_{n_i}$ and $\theta_{n_j}$. Up to smooth prefactors, this yields
\begin{equation}
|\tilde{\Psi}(x_1,\dots,x_N)|^2
\propto
\prod_{i<j} (x_i - x_j)^2.
\end{equation}
Thus, in the continuum limit, the squared many-body wavefunction acquires the universal Vandermonde form characteristic of free fermions \cite{C0}, reflecting the effective repulsion induced by the Pauli exclusion principle \cite{C0}. Boundary-dependent prefactors (such as $\prod_i \sin\theta_{n_i}$) contribute only smooth envelope corrections and do not affect the large-scale structure of the density.
Including the non-Hermitian factor, the full many-body probability density reads
\begin{equation}
|\Psi(x_1,\dots,x_N)|^2
\propto
\exp\left(-2\sum_i G(x_i)\right)
\prod_{i<j} (x_i - x_j)^2,
\end{equation}
where $G(x)$ is the cumulative non-Hermitian gauge field. This expression can be interpreted as a classical Gibbs distribution \cite{C0,C0bis,C1,C2,C3}
\begin{equation}
P(\{x_i\}) \propto e^{-\mathcal{H}(\{x_i\})},
\end{equation}
with effective Hamiltonian
\begin{equation}
\mathcal{H} =
-2 \sum_{i<j} \log |x_i - x_j|
+ 2 \sum_i G(x_i).
\end{equation}
In the thermodynamic limit, this Coulomb-gas representation yields a continuum description in terms of a coarse-grained density $\rho(x)$ satisfying
\begin{equation}
\int dx\, \rho(x) = \nu.
\end{equation}
The corresponding free-energy functional reads
\begin{equation}
\mathcal{F}[\rho] =
 \int dx\, \rho(x) G(x)
- \int dx\,dy\, \rho(x)\rho(y)\log|x-y|.
\end{equation}
The equilibrium density is obtained by minimizing $\mathcal{F}[\rho]$ under the normalization constraint. Introducing a Lagrange multiplier $\lambda$, one minimizes
\begin{equation}
\mathcal{F}[\rho] - \lambda \left( \int dx\, \rho(x) - \nu \right).
\end{equation}
The saddle-point condition is
\begin{equation}
G(x) = 2 \int dy\, \rho(y)\log|x-y| + \lambda.
\end{equation}

\medskip

\noindent
For smooth potentials $G(x)$, the saddle-point equation admits solutions with a single connected support, corresponding to a single real-space Fermi surface. In contrast, when $G(x)$ is stochastic and inherits the fluctuations of a random walk, the effective potential develops multiple local extrema. As a consequence, the support of $\rho(x)$ generically splits into several disconnected intervals, corresponding to fragmented real-space Fermi domains. The number of such intervals is controlled by the statistics of the extrema of $G(x)$, leading to a subextensive scaling of the number of Fermi surface segments.

To illustrate this mechanism, it is useful to consider a simple solvable case with a smooth confining potential. For a quadratic potential of the form
$G(x) = \kappa (x-1/2)^2$,
the Coulomb gas admits a solution with a single connected support centered at the minimum of the potential. The corresponding equilibrium density takes the Wigner semicircle-like form \cite{C0,C0bis,C1,C2,C3},
\begin{equation}
\rho(x) \simeq \frac{1}{\pi} \sqrt{2\mu - \kappa (x-1/2)^2},
\end{equation}
with support fixed by normalization and constrained to lie within the interval $(0,1)$.
When $G(x)$ is generated by a random walk, it develops fluctuations on all scales and contains multiple local extrema. In contrast to smooth potentials, the local structure of $G(x)$ near its extrema is typically non-quadratic and is more accurately described by piecewise linear behavior of the form $G(x)\sim \kappa |x-x_{\min}|$ [see e.g. Fig.5(b)]. As a result, the associated density profiles near each minimum do not follow the semicircular form, but instead exhibit flatter, more square-like shapes, as observed in the numerical simulations. 
For sufficiently strong disorder, the density can thus be viewed as a
collection of weakly overlapping droplets localized near the relevant
large-scale minima of $G(x)$, whose shape is controlled by the local
structure of the stochastic potential. {\color{black} The number of ordinary local extrema
of $G(x)$ is extensive and does not directly determine the number of density
domains. Rather, the Pauli-induced logarithmic repulsion coarse-grains the
random-walk landscape, making the density sensitive to its hierarchy of
deep, record-like excursions. Their universal first-passage/record statistics
provide the natural $L^{1/2}$ scale \cite{MajumdarZiff2008,NeveuPitman1989},
while the robust signature of fragmentation is the resulting subextensive
number of spatially separated density domains.}

\end{document}